\documentclass[12pt,a4paper,english,superscriptaddress,aps,nofootinbib]{revtex4}
\usepackage[utf8]{inputenc}
\usepackage[T1]{fontenc}
\usepackage{amsmath,amssymb,graphicx}
\makeatletter
\usepackage{babel}
\usepackage[active]{srcltx}
\usepackage{graphicx,color}
\usepackage{subfigure}
\usepackage{changebar}

\usepackage{hyperref}
\hypersetup{colorlinks=true,urlcolor= blue,citecolor=blue,linkcolor= blue}
\usepackage{braket}
\usepackage{dsfont}
\usepackage{mathtools}
\usepackage{slashed}
\usepackage{empheq}
\usepackage{tikz}
\usepackage{multirow}
\usetikzlibrary{decorations.pathmorphing}
\usepackage{graphicx}
\usepackage{amsfonts}
\usepackage{amsmath}
\newcommand{\be}{\begin{equation}}
	\newcommand{\ee}{\end{equation}}
\newcommand{\bea}{\begin{eqnarray}}
	\newcommand{\eea}{\end{eqnarray}}

\begin{document}

\title{Geometrically expanding the BPS vortex of a non-canonical multi-field theory}

\author{F. C. E. Lima}
\email[]{E-mail: cleiton.estevao@ufabc.edu.br}
\affiliation{Centro de Matématica, Computação e Cognição (CMCC), Universidade Federal do ABC (UFABC), Av. dos Estados 5001, CEP 09210-580, Santo Andr\'{e}, S\~{a}o Paulo, Brazil.}

\begin{abstract}
\noindent\textbf{Abstract:} Considering a non-canonical multi-field theory, we propose a mechanism of geometric expansion for Abrikosov-Nielsen–Olesen (ANO) vortices. One builds the general setup adopting a non-canonical O(3)-sigma model, non-minimally coupled through an anomalous magnetic dipole interaction to a gauge and real scalar field. By embracing a hyperbolic non-canonical extension, one finds that self-dual vortex configurations undergo geometric expansion, deforming ANO-like vortices into disk-like structures, mitigating their energy, and generating concentric energy rings around the core. Furthermore, these vortices exhibit quantized magnetic flux and degenerate solutions. Finally, we highlight that the analyzed vortices carry energy below the corresponding magnetic flux while preserving a singularity at their origin.

\end{abstract}

\maketitle

\thispagestyle{empty}

\newpage

\section{Introduction}

The study of topological solutions in field theories has attracted considerable attention within the academic community, both for its theoretical significance \cite{Adam1,Adam2,Adam3,Dmitri,Adalto} and for its potential applications \cite{Chen,Wang}. Within these solutions, magnetic vortices stand out as localized structures associated with magnetic flux and stable energy states \cite{Bogdanov,Kobler,Nagaosa}. The Abrikosov-Nielsen-Olesen (ANO) provides us a classical example of vortices, offering a description of superconductors \cite{Abrikosov,Nielsen}. Accordingly, these structures have been the subject of extensive research within several fields of physics, from string theory \cite{Ovrut,Horn,Shiftman} to condensed matter \cite{Palacios,Yi}.

In broad terms, these vortices emerge as soliton-like solutions in gauge models with spontaneous symmetry breaking, exhibiting physical properties of magnetic flux confinement and topological stability \cite{Nitta,Chagoya}. Furthermore, an intrinsic feature of the system is the persistent presence of a winding number, safeguarding the vortex against smooth deformations \cite{Rajaraman,Manton}. Besides, it is worth noting that these structures are related to spontaneous symmetry breaking, under this framework, when the theory exhibits the Bogomol’nyi–Prasad–Sommerfield (BPS) property, it induces an energetic rearrangement of the fields, defining a lower energy bound for the BPS vortices \cite{Bogomolnyi,PSommerfield}. Within this framework, the equations of motion, originally of second order, can be reduced to a system of first-order differential equations, which not only facilitates the analysis but also highlights the relation between the minimal energy of field configurations and their topological charge. 

Recent literature highlights the need to extend traditional models to account for non-canonical dynamics and new couplings that emerge in more complex scenarios \cite{Schoer,FCELima1,FCELima2,FCELima3}. In magnetic materials with exotic crystal structures (and in cosmological systems descriptions), nontrivial effects related to nonlinear couplings and Anomalous Magnetic Dipole Moments (AMDM) \cite{Cunha} can substantially modify the behavior of topological structures, allowing, for instance, the emergence of internal structures \cite{Igor}. Consequently, such extensions enable the appearance of new phenomena, including modifications of energy profiles, the creation and decay of vacua \cite{Igor,FLima1}, among others.

Within this setting, the present project proposes the construction of a geometric expansion mechanism for ANO-like vortices, based on an extension of the O(3)-sigma model, incorporating an AMDM coupling to a gauge field and an additional scalar field. The innovative aspect of this approach consists of the implementation of an AMDM coupling \cite{Cunha,Igor}, combined with a hyperbolic-like generalizing function \cite{FLima1} that connects the field sectors. This combination allows for the emergence of self-dual (BPS) vortices that expand geometrically, transforming the characteristic point-like core of Abrikosov vortices into disk-shaped structures, with energy profiles exhibiting concentric ring patterns.

From a physical viewpoint, this geometric deformation raises several questions, viz., do the vortices maintain quantized magnetic flux, thereby evidencing the preservation of their topological nature even under the geometric expansion mechanism? Does the expansion process reduce the total energy of the system relative to the associated flux? Are the magnetic vortices emerging in this context energetically more favorable? These are among the key questions that our study aims to address throughout its development.

Beyond their conceptual significance, there is also an applied motivation. Once structures similar to those investigated here have parallels in type-II superconductor vortices, which play a fundamental role in advanced technologies such as superconducting magnets \cite{Wilson,Kleiner}, high-precision sensors (SQUIDs) \cite{Hari}, and medical detection devices \cite{Wilson,Kleiner,Hari}; within this perspective, one can understand how vortices deform, expand, and redistribute energy in a controlled manner, which may contribute to the development of new theoretical models and to practical applications \cite{Lima2}.

Therefore, the focus of this work is to study the formation, properties, and physical implications of geometrically expanded vortices in a non-canonical multi-field theory, exploring both the analytical and numerical solutions of the vortex configurations. Employing this approach, we aim to provide a solid theoretical foundation for understanding how the geometric expansion mechanism and its repercussions on the physical aspects of ANO-like vortices by offering new perspectives for the study of these structures. 

We outlined our work into four sections. In Sect. II, we present the general setup for describing magnetic vortices in a non-canonical multi-field theory subject to an arbitrary potential. This section also discusses the BPS formulation and the profiles of the sigma, gauge, and scalar fields in terms of the field variables, highlighting the physical properties of the vortices. Posteriorly, one examines the physical aspects of these vortices, including their analytical and numerical solutions in Sect. III. Finally, announced our findings in Sect. IV.

\section{The general framework for vortex structures}

To embark on our study, let us consider a non-linear and non-canonical O(3)-sigma model, coupled in a non-minimal way to the electromagnetic sector and to an additional scalar field. One justifies this choice by the fact that, in principle, the solitons can arise in crystals governed by non-canonical dynamics and subject to contributions arising from the Anomalous Magnetic Dipole Moment (AMDM) \cite{Canfora}. This possibility arises from the O(3)-sigma model's ability to depict solitonic profiles that enable the emergence of electromagnetic topological structures, such as electromagnetic vortices \cite{Lima1}. Furthermore, in constructing our model, we assume that the appearance of topological vortex structures is, in part, conditioned by the contribution of the AMDM \cite{Chaudhuri}. Consequently, the physical and energetic profiles of the topological structures are susceptible to modifications due to the interaction with this term. Taking into consideration these aspects, we adopt the action\footnote{In this work, we consider a three-dimensional flat spacetime with the metric $\eta_{\mu\nu}=\text{diag}(+1,-1,-1)$.}
\begin{align}\label{Eq1}
	S=\int\,d^3x\, \left[\frac{\mathcal{F}(\psi)}{2}\nabla_\mu\Phi\cdot\nabla^\mu\Phi-\frac{1}{4}F_{\mu\nu}F^{\mu\nu}+\frac{1}{2}\partial_{\mu}\psi\,\partial^{\mu}\psi-V[\Phi(\phi_3),\psi]\right].
\end{align}
Within this framework, $\Phi$ is the O(3)-sigma field \footnote{The non-linear O(3)-sigma field is a three-dimensional vector field $\Phi = (\phi_1(x_\mu), \phi_2(x_\mu), \phi_3(x_\mu))$, defined in the three-dimensional spacetime and subject to the constraint $\Phi \cdot \Phi = \phi_1^2 + \phi_2^2 + \phi_3^2 = 1$.}, $\psi$ is a real scalar field, $F_{\mu\nu}$ is the electromagnetic tensor \footnote{One defines the electromagnetic tensor as $F_{\mu\nu} = \partial_\mu A_\nu - \partial_\nu A_\mu$, where $A_\mu$ is the gauge field.}, and $V[\Phi(\phi_3),\psi]$ is the potential responsible for spontaneous symmetry breaking in the topological sectors of the $\Phi$ and $\psi$ fields. Additionally, we introduce the generalizing function $\mathcal{F}(\psi)$, which connects the sectors of the scalar field and the sigma field $\Phi$, thereby enabling the emergence of spontaneous symmetry breaking in both topological sectors. From a physical perspective, this function effectively modifies the potential, potentially inducing new phenomena, e.g., the emergence of new vacua, shifts in the positions of existing vacua, and the possible decay of false vacuum states.

Under these circumstances, the covariant derivative is
\begin{align}\label{Eq2}
	\nabla_\mu\Phi=\partial_\mu\Phi+\left(e A_\mu+\frac{g}{2}\varepsilon_{\mu\nu\lambda}F^{\nu\lambda}\right)(\hat{n}_3\times\Phi),
\end{align}
where the term $\frac{g}{2}\,\varepsilon_{\mu\nu\lambda}F^{\nu\lambda}$ is the contribution of the AMDM, $\hat{n}_3 = (0,0,1)$ is a unit vector in the direction of the $\phi_3$ of the sigma field. Furthermore, $\partial_\mu \Phi$ is the ordinary covariant derivative, and $e$ the particle's charge.

%Henceforth, let us proceed with the investigation of the equations of motion concerning the action \eqref{Eq1}. Toward this purpose, it is necessary to vary the action concerning the gauge field $A_\mu$, the sigma field $\Phi$, and the scalar field $\psi$. Particularly, the variation of the action \eqref{Eq1} concerning the gauge field yields the expression 
%\begin{align}\label{Eq3}
%\partial_\nu[\mathcal{F}\,g\,\varepsilon_{\mu}\,^{\nu\lambda}(\Phi\times\nabla^\mu\Phi)\cdot\hat{n}_3-   ]F^{\nu\lambda}]=j^\lambda,
%\end{align}
%in which $j^\lambda$ corresponds to the current \footnote{Naturally, the zeroth component is the Gauss's law \cite{Rubakov}.}. Mathematically, one defines this current as 
%\begin{align}\label{Eq4}
%	j^\lambda=-2\,e\,\mathcal{F}(\Phi\times\nabla^\lambda\Phi)\cdot\hat{n}_3.
%\end{align}

%Meanwhile, the variation of the action \eqref{Eq1} concerning the sigma field leads to the equation of motion, viz.,
%\begin{align}\label{Eq5}%
%	\nabla_\nu(\mathcal{F}\,\nabla^\nu\Phi)=-V_{\Phi},
%\end{align}
%with $V_{\Phi}=\frac{\partial V}{\partial \Phi}$.

%Finally, by varying action \eqref{Eq1} with respect to the scalar field $\psi$, one arrives at the expression
%\begin{align}\label{Eq6}
%	\partial_\mu\partial^\mu\psi-\frac{\mathcal{F}_\psi}{2}\nabla_\mu\Phi\cdot\nabla^\mu\Phi+V_\psi=0,
%\end{align}
%to which $\mathcal{F}_\psi=\frac{d\mathcal{F}}{d\psi}$ and $V_\psi=\frac{\partial V}{\partial \psi}$.

Within this framework, the energy\footnote{One defines the energy as the $d$-dimensional integral (in this case, $d=2$) of the 00-component of the energy–momentum tensor. It is worth noting that the energy–momentum tensor is given by
$T^{\mu}\,_{\nu} = \frac{\partial \mathcal{L}}{\partial (\partial_\mu \varphi^i)} \, \partial_\nu \varphi^i - \delta^{\mu}\,_{\nu} \, \mathcal{L}$, where $\varphi^i$ is the $i$-th field and $\mathcal{L}$ the Lagrangian density.} of the vortices is
\begin{align}\label{Eq7}
	\text{E}=\int\,d^2x\, \left[\frac{\mathcal{F}}{2}(\nabla_i\Phi)^2+\frac{1}{2}F_{ij}F^{ij}+\frac{1}{2}(\partial_i\psi)^2+V[\Phi(\phi_3),\psi]\right].
\end{align}
Naturally, the energy can be summarized through the equations of motion if the BPS property is omnipresent. %In this case, the equations of motion [\eqref{Eq3}, \eqref{Eq5}, and \eqref{Eq6}] can be reduced to a system of coupled first-order equations.

\subsection{The BPS approach}

To verify the BPS property, we shall proceed by considering the energy from the vortex configurations. In this framework, the energy announced in Eq. \eqref{Eq7} boils down to
\begin{align}\label{Eq8}
	\text{E}=\frac{1}{2}\,\int\,d^2x\, \left[\mathcal{F}(\nabla_i\Phi)^2+(F_{ij})^2+(\partial_i\psi)^2+2V[\Phi(\phi_3),\psi]\right].
\end{align}

As reported by Ghosh et al. \cite{PGhosh1,PGhosh2}, the sigma-$O(3)$ model admits vortex configurations described either by topological and/or non-topological field profiles\footnote{Topological vortices are localized field solutions. These configurations are stable, with their stability determined by the topology of the target space. Consequently, such configurations cannot be continuously deformed away without introducing discontinuities, meaning they are topologically protected. Naturally, this leads to three fundamental features of these structures: they arise from a spontaneous symmetry breaking concerning a nontrivial vacuum manifold, in which ensure their stability by the winding number (topological invariant). \cite{Rajaraman,Manton,Shnir}.}. Therefore, it is essential to employ the BPS approach\footnote{The BPS approach becomes particularly relevant once it allows the study of vortex structures by simplifying their equations of motion, exposing the self-dual equations. In this procedure, the second-order equations boil down to first-order ones by simplifying energy when $\text{E}=\text{E}_{\text{BPS}}$. That facilitates the derivation of stable analytical solutions with minimal energy, relating the energy to the topological charge, while also enabling the identification of physically relevant solutions even in the absence of topological protection (non-topological cases). Furthermore, the BPS condition is associated with properties of supersymmetry, stability, and the classification of solutions, making it a central tool in the analysis of these solitons. For further details on BPS configurations, see Ref. \cite{Rajaraman,Manton,Shnir,Bogomolnyi,PSommerfield,CAdam}.} to analyze the self-dual equation concerning the action \eqref{Eq1}. Accordingly, to investigate the BPS property, we write the vortex energy as
\begin{align}\nonumber \label{Eq9}
	\text{E}=&\int\, d^2x\, \bigg[\frac{\mathcal{F}}{2}\left(\nabla_i\Phi\mp\frac{1}{\sqrt{\mathcal{F}}}\varepsilon_{ij}\,\Phi\times\nabla_{j}\Phi\right)^2+\frac{1}{2}(F_{ij}\pm\sqrt{\tilde{W}})^2+\frac{1}{2}\left(\partial_i\psi\mp\frac{W_\psi}{r}\right)^2+V+\\
\mp&\varepsilon_{ij}\Phi\cdot(\nabla_i\Phi\times\nabla_j\Phi)\mp F_{ij}\sqrt{\tilde{W}}\pm\frac{1}{r}W_{\psi}\partial_i\psi-\tilde{W}-\frac{W_{\psi}^{2}}{2r^2}\bigg].
\end{align}
Within this framework, let us introduce the auxiliary functions $\tilde{W}(\Phi)=\tilde{W}[\Phi(\phi_3)]$ and $W = W(\psi)$. These functions are the superpotentials of the sigma ($\Phi$) and scalar ($\psi$) fields. The use of these functions is relevant, as it simplifies the analysis of the BPS property by establishing a direct relation between the superpotentials ($W$ and $\tilde{W}$) and the potential ($V$) and total energy ($\text{E}$) \footnote{For further details on this approach, see Ref. \cite{CAdam2}.}.

By adopting the superpotentials $\tilde{W}$ and $W$, the field energy can be reformulated as
\begin{align}\nonumber \label{Eq10}
	\text{E}=&\int\, d^2x\, \bigg[\frac{\mathcal{F}}{2}\left(\nabla_i\Phi\mp\frac{1}{\sqrt{\mathcal{F}}}\varepsilon_{ij}\,\Phi\times\nabla_{j}\Phi\right)^2+\frac{1}{2}(F_{ij}\pm\sqrt{\tilde{W}})^2+\frac{1}{2}\left(\partial_i\psi\mp\frac{W_\psi}{r}\right)^2\mp F_{ij}\sqrt{\tilde{W}}\\
\mp&\varepsilon_{ij}\Phi\cdot(\nabla_i\Phi\times\nabla_j\Phi)\pm\frac{1}{r}W_{\psi}\partial_i\psi\bigg],
\end{align}
allowing us to conclude that, for the model to have BPS property, the potential $V[\Phi(\phi_3,\psi)]$ must be
\begin{align}\label{Eq11}
	V[\Phi(\phi_3),\psi]=\tilde{W}+\frac{W_{\psi}^2}{2r^2},
\end{align}
which leads us to 
\begin{align}\label{Eq12}
	\text{E}=&\int\, d^2x\, \left[\frac{\mathcal{F}}{2}\left(\nabla_i\Phi\mp\frac{1}{\sqrt{\mathcal{F}}}\varepsilon_{ij}\,\Phi\times\nabla_{j}\Phi\right)^2+\frac{1}{2}(F_{ij}\pm\sqrt{\tilde{W}})+\frac{1}{2}\left(\partial_i\psi\mp\frac{W_\psi}{r}\right)^2\right]+\text{E}_{\text{BPS}}.
\end{align}
Under these circumstances, $\text{E}_{\text{BPS}}$ is the BPS energy. Mathematically, the BPS energy is
\begin{align}\label{Eq13}
	\text{E}_{\text{BPS}}=\mp&\int\, d^2x\,\left[\varepsilon_{ij}\Phi\cdot(\nabla_i\Phi\times\nabla_j\Phi)+ F_{ij}\sqrt{\tilde{W}}-\frac{1}{r}W_{\psi}\partial_i\psi\right].
\end{align}
Here, we highlight that the energy is bounded from below, i.e., $\text{E} \geq \text{E}_{\text{BPS}}$.

Therefore, by saturating the energy of the vortex configuration (i.e., $\text{E} = \text{E}_{\text{BPS}} $), one obtains the self-dual equations (or BPS equations), viz.:
\begin{align}\label{Eq14}
    \nabla_i\Phi=\pm\frac{1}{\sqrt{\mathcal{F}}}\varepsilon_{ij}\,\Phi\times\nabla_j\Phi, \hspace{1cm} F_{ij}=\mp \sqrt{\tilde{W}}, \hspace{1cm} \text{and} \hspace{1cm} \partial_i\psi=\pm\frac{W_\psi}{r}.
\end{align}

\subsection{The magnetic vortex configurations}

Hereafter, let us specialize our model to the case of vortex structures from the O(3)-sigma model coupled to the gauge and scalar fields. To describe the topological sector of the O(3)-sigma model, the sigma field assumes the profile\footnote{Note that this profile ensures the validity of the constraint $\Phi\cdot\Phi=1$ \cite{Rajaraman}.}
\begin{align}\label{Eq15}
	\Phi=\begin{pmatrix}
		\sin f(r)\cos N\Theta\\
		\sin f(r)\sin N\Theta\\
		\cos f(r)
	\end{pmatrix}.
\end{align}
Meanwhile, for the generation of vortex lines \cite{Nielsen}, the gauge field takes the form 
\begin{align}\label{Eq16}
    \textbf{A}=-\frac{N}{e r}a(r)\,\hat{\text{e}}_{_\Theta}.
\end{align}
Here, $N$ is the winding number, $f(r)$ is the field variable concerning the sigma field, $\Theta$ is the angular variable, and $r$ is the radial variable. 

For simplicity, and to ensure that only soliton-like configurations at the topological sector from the scalar field affect the sigma sector, let us adopt the premise in which the scalar field $\psi$ depends solely on the radial coordinate. Thereby, the field $\psi$ is
\begin{align}\label{Eq17}
    \psi=\psi(r).
\end{align}

It becomes convenient to examine the vortex configurations in a static regime. In this framework, analyzing the Gauss law%\footnote{The Gauss law corresponds to the analysis of the zeroth component of the equations \eqref{Eq3} and \eqref{Eq4}.}
, one obtains
\begin{align}\label{Eq18}
    j^0=& \partial_\nu[\mathcal{F}\,g\,\varepsilon_{\mu}\,^{\nu\, 0}(\Phi\times\nabla^\mu\Phi)\cdot\hat{n}_3-F^{\nu\, 0}]=0,
\end{align}
which allows us to conclude that $A^0=0$. Therefore, only purely magnetic vortices emerge in the theory. 

Once the nature of the vortices is purely magnetic, their magnetic flux ($\varphi_{\text{flux}}$) will be 
\begin{align}\label{Eq19}
    \varphi_{\text{flux}}=-\int_{\Theta=0}^{\Theta=2\pi}\,\int_{\bar{r}=0}^{\bar{r}=\infty}\, \bar{r}\,d\bar{r}\,d\Theta\, \left(\frac{N\,a'(\bar{r})}{e\bar{r}}\right)=\frac{2\pi N}{e}[a(0)-a(\infty)].
\end{align}

Furthermore, the BPS energy \eqref{Eq13} from the vortices boils down to
\begin{align}\label{Eq20}
	\text{E}_{\text{BPS}}=\pm\int\,d^2x\left[\frac{f'}{r}N(a-1)\sin f-\frac{N a'}{r}\sqrt{\tilde{W}}+\frac{1}{r}\frac{\partial W}{\partial x}\right],
\end{align}
which prompts us to constrain $\tilde{W}=\cos^2 f$ to ensure the BPS property in our model. Furthermore,  the contribution of the electromagnetic tensor boils down to $F_{ij}=-F_{ji}=F_{12}=-\frac{Na'}{er}$ \footnote{In this article, we will adopt natural units, i.e., $\hbar=e=c=1$.}.

For the theory to have the BPS property, the constraint $\tilde{W} = \cos^2 f$ must be satisfied. Thus, one concludes that the BPS energy density reduces to
\begin{align}\label{Eq21}
\text{E}_{\text{BPS}}=\pm\int_{\bar{\Theta}=0}^{\bar{\Theta}=2\pi}\int_{\bar{r}=0}^{\bar{r}=\infty}\, \bar{r}\,d\bar{r}\,d\bar{\Theta}\left[-\frac{N}{\bar{r}}\frac{d}{d\bar{r}}[(a-1)\cos f]+\frac{1}{\bar{r}}\frac{d W}{d \bar{r}}\right],
\end{align}
allowing us to arrive at
\begin{align}\label{Eq22}
	\text{E}_{\text{BPS}}=\pm\,[W-N(a-1)\sqrt{\tilde{W}}]\bigg\vert_{r=0}^{r=\infty}.
\end{align}

Within the BPS domain, the self-dual equations in terms of the field variable are:
\begin{align}\label{Eq23}
	&f'=\pm\frac{1}{\sqrt{\mathcal{F}}}\frac{N}{r}(a-1)\sin f, \hspace{1cm} a'=\pm\frac{\sqrt{2}r}{N}\cos f, \hspace{1cm} \text{and} \hspace{1cm} \psi'=\pm\frac{W_\psi}{r}.
\end{align}

\section{Vortex solution of the multi-field theory}

\subsection{The boundary condition}

To pursue this research, one defines the topological boundary conditions for the field variables $f(r)$ and $a(r)$ as\footnote{A topological boundary conditions require that, at the limits of space, e.g., $r \to 0$ and $r \to \pm \infty$, the field takes values corresponding to distinct minima of the potential. These conditions allow us to classify solutions into topological sectors associated with the structure of the vacuum manifold, thereby conferring stability to nontrivial configurations. \cite{Rajaraman, Manton}.}
\begin{align}\label{Eq24}
    f(0)=0, \hspace{1cm} f(\infty)=\pi, \hspace{1cm} a(0)=0, \hspace{1cm} \text{and} \hspace{1cm} a(\infty)=-\beta.
\end{align}
Here, the parameter $\beta\in\,\mathds{R}$ corresponds to the contribution of the vortex's magnetic flux far from its core.

Besides, taking into account that $r \in [0, +\infty]$, the topological structures concerning the scalar field topological sector must satisfy the conditions
\begin{align}\label{Eq25}
    \psi(0)=-\nu \hspace{1cm} \text{and} \hspace{1cm} \psi(\infty)=+\nu,
\end{align}
where $\nu$ is the Vacuum Expectation Value (VEV).

\subsection{Physical aspects of magnetic vortices}

Once the topological boundaries are well defined [Eq. \eqref{Eq24}], the vortex magnetic flux is obtained by adopting Eq. \eqref{Eq19}, which allows us to conclude that
\begin{align}\label{Eq26}
    \varphi_{\text{flux}}=\frac{2\pi N \beta}{e},
\end{align}
where $N\in\mathds{Z}$. Thus, one notes that the vortex has quantized magnetic flux, i.e.,
\begin{align}
    \label{Eq27}
    \varphi_{\text{flux}}=\frac{2\pi \beta}{e},\frac{4\pi \beta}{e},\frac{6\pi \beta}{e},\frac{8\pi \beta}{e},\frac{10\pi \beta}{e},\,\dots \,.
\end{align}

Meanwhile, assuming the vortex energy given in Eq. \eqref{Eq22}, we obtain
\begin{align}\label{Eq28}
    \text{E}_{\text{BPS}}=\pm2\pi[\Delta W+N \Delta \overline{W}+N_\beta\,\tilde{W}_{\infty}^{1/2}],    
\end{align}
such that $N_\beta=N\beta$, $\Delta W = W_\infty-W_0$, and $\Delta \overline{W}=\tilde{W}_{\infty}^{1/2} - \tilde{W}_0^{1/2}$. Thus, the energy profile highlights the topological nature of the magnetic vortices. It is noteworthy that the energy contributions of the $\Phi$ and $\psi$ fields are, respectively, $\text{E}_{\text{BPS}}^{(\Phi)}=\pm 2\pi[N\,\Delta\overline{W}+N\beta\,\tilde{W}{\infty}^{1/2}]$ and $\text{E}_{\text{BPS}}^{(\psi)}=\pm 2\pi\Delta W$. Therefore, Eq. \eqref{Eq28} can be reduced to $\text{E}_{\text{BPS}}=\text{E}_{\text{BPS}}^{(\Phi)}+\text{E}_{\text{BPS}}^{(\psi)}$.

\subsection{Topological solution of the scalar field}

To move forward, let us adopt a $\varphi^4$-like superpotential in the scalar field topological sector. Under these circumstances, the superpotential $W$ takes the form\footnote{We are motivated to adopt a $\varphi^4$-like theory, once it plays a prominent role in theoretical physics as the simplest interacting scalar field model, serving as a foundation for the study of essential concepts in quantum field theory \cite{Ryder}. Beyond its pedagogical value, we find several applications in various branches of physics, such as the description of phase transitions in critical systems \cite{Radescu}, the study of classical solutions (topological and non-topological solitons) \cite{Lima1,Lima2}, among others.}
\begin{align}\label{Eq29}
	W(\psi)=\sqrt{\lambda}\left(\nu^2\psi-\frac{\psi^3}{3}\right),
\end{align}
allowing us to write the equation of motion for the scalar field $\psi$ as
\begin{align}\label{Eq30}
	\psi'=\pm\frac{\sqrt{\lambda}}{r}(\nu^2-\psi^2).
\end{align}
In this framework, the prime notation describes the derivative with respect to the radial variable $r$.

\begin{figure}[!ht]
    \centering
    \includegraphics[width=8cm,height=7cm]{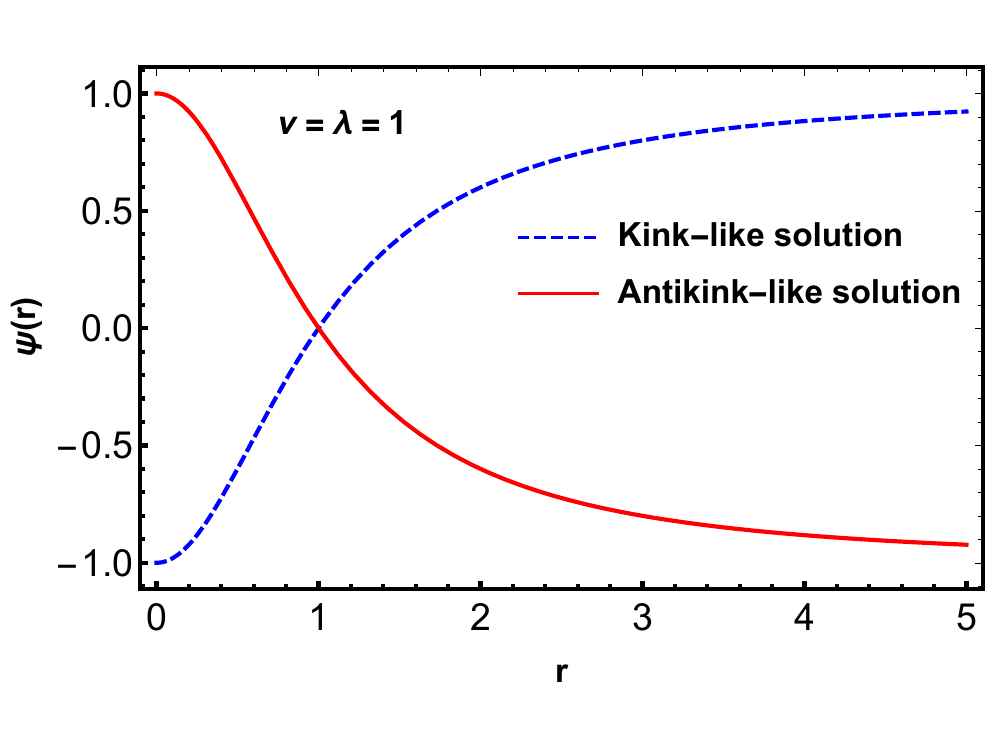}
    \caption{Scalar field solution $\psi(r)$ vs. $r$ adopting the parameters $\lambda=\nu=1$.}
    \label{Fig1}
\end{figure}

By examining the solution of Eq. \eqref{Eq30} together with the boundary conditions \eqref{Eq25}, one obtains
\begin{align}\label{Eq31}
	\psi(r)=\pm\nu\,\tanh\left[\eta\,\text{ln}\left(r\right)\right],
\end{align}
where $\eta = \nu \sqrt{\lambda}$. Therefore, we conclude that the topological solutions of the scalar field are kink-like (positive sign) and antikink-like solutions (negative sign). Details of these solutions, announced in Eq. \eqref{Eq31}, are shown in Fig. \ref{Fig1}.

\subsection{The sigma field solution}

\subsubsection{Computational approach}\label{sec2d1}

To investigate the solutions of magnetic vortices, it is necessary to examine the solutions that simultaneously satisfy the system of equations \eqref{Eq23} subject to the boundary conditions \eqref{Eq24}. In this case, the system of equations \eqref{Eq23} admits only numerical solutions. For this purpose, we employ the Runge–Kutta numerical method, which proves to be particularly efficient for analyzing equations of type
\begin{align}\label{Eq32}
    \frac{dg_1}{d\xi}=h_1(g_1,g_2;\xi) \hspace{1cm} \text{and} \hspace{1cm} \frac{dg_2}{d\xi}=h_2(g_1,g_2;\xi),
\end{align}
with initial conditions $g_1(\xi_0)=g_{1,0}$ and $g_2(\xi_0)=g_{2,0}$. For solving our numerical problem, becomes necessary to adopt a finite range, i.e., $\xi \in [\xi_1,\xi_2]$, with discretization in steps of size $h$.

Thus, the premise for obtaining the numerical solutions consist in the application of the fourth-order Runge–Kutta method simultaneously to all variables of the system\footnote{In this case, the independent variable is $\xi$, and the dependent variables are $g_1$ and $g_2$.} \eqref{Eq32} for each time step $\xi_n \to \xi_{n+1} = \xi_n + h$, which allows us to numerically compute the coefficients: 
\begin{align}\nonumber
    & k_{1,1}=h_1(\xi_n, g_{1,n}, g_{2,n});\\ \nonumber
    & k_{1,2}=h_2(\xi_n, g_{1,n}, g_{2,n});\\ \nonumber
    & k_{2,1}=h_1\left(\xi_n+\frac{h}{2}, g_{1,n}+\frac{h}{2}k_{1,1}, g_{2,n}+\frac{h}{2}k_{1,2}\right);\\ \label{Eq33}
    & k_{2,2}=h_2\left(\xi_n+\frac{h}{2}, g_{1,n}+\frac{h}{2}k_{1,1}, g_{2,n}+\frac{h}{2}k_{1,2}\right);\\ \nonumber
    & k_{3,1}=h_1\left(\xi_n+\frac{h}{2}, g_{1,n}+\frac{h}{2}k_{2,1}, g_{2,n}+\frac{h}{2}k_{2,2}\right);\\ \nonumber
    & k_{3,2}=h_2\left(\xi_n+\frac{h}{2}, g_{1,n}+\frac{h}{2}k_{2,1}, g_{2,n}+\frac{h}{2}k_{2,2}\right);\\ \nonumber
    & k_{4,1}=h_1(\xi_n+h\, g_{1,n}+h\, k_{3,1}, g_{2,n}+h\, k_{3,2});\\ \nonumber
    & k_{4,2}=h_1(\xi_n+h\, g_{1,n}+h\, k_{3,1}, g_{2,n}+h\, k_{3,2}).
\end{align}
Therefore, using the coefficients announced in Eq. \eqref{Eq33}, one finds the $n$-th solutions, namely, 
\begin{align}
    \label{Eq41}
    g_{1,n+1}=g_{1,n}+\frac{h}{6}(k_{1,1}+2k_{2,1}+2k_{3,1}+k_{4,1})
\end{align}
and
\begin{align}
    \label{Eq42}
    g_{2,n+1}= g_{2,n}+\frac{h}{6}(k_{1,2}+2k_{2,2}+2k_{3,2}+k_{4,2}).
\end{align}
For our simulations, we discretized the radial variable $r$ of the vortex [see Eqs. \eqref{Eq23}] over a well-defined and discrete range. We performed this discretization by subdividing the range into ten thousand discrete points. Thus, the solutions of the differential equation are investigated within the interval $[0, 10]$, using the discretization $r_1, r_2, \dots, r_{10000}$, where $r_1 = 0$ and $r_{10000} = 10$.

To summarize, we estimate the values of the field variables, i.e., $(r_n, f_n)$ and $(r_n, a_n)$ [Eq. \eqref{Eq23}], over the discrete range $[0,10]$ with steps of $h=10^{-10}$. We then apply numerical interpolation to construct continuous numerical solutions for the field variables $f(r)$ and $a(r)$. In general, we employ low-degree polynomial functions to ensure smooth transitions between discrete points and the existence of derivatives. For further details, see Refs. \cite{Burden,Butcher,Atkinson}.

\subsubsection{Numerical solutions of the vortices: topological sector of the sigma field}

Finally, we are ready to examine the numerical solutions of the remaining field variables that describe the magnetic vortices in the multi-field theory \eqref{Eq1}. To move forward with our study, let us adopt a profile for the generalizing function\footnote{This hyperbolic profile of the generalizing function has been widely studied in topological theories, e.g., to examine the impact of non-canonical theories on domain walls \cite{Lima2}. However, this generalizing function becomes particularly interesting in the BPS limit once the hyperbolic non-canonical theory yields new $\phi^4$-like hyperbolic theories in both sectors, i.e., the scalar field and the sigma field. Moreover, this hyperbolically modified theory is relevant due to its connection with the description of ferroelectric and ferromagnetic materials \cite{Simas3}. Additionally, hyperbolic generalized functions effectively describe the fragility and increased stiffness of crystal lattices in various materials, such as ferroelectric materials \cite{Lima2}. Hyperbolic models also arise as extensions of the Calogero model, allowing for the formation of multiple trapped solitons \cite{Calogero,Gon}.}, viz., 
\begin{align}\label{Eq43}
	\mathcal{F}(\psi)=\sinh[q\, \psi^2]^{-\frac{1}{2z+1}},
\end{align} 
where the parameter $q$ has the canonical dimension corresponding to the inverse square of the scalar field $\psi$, i.e., $[q^{-2}] \equiv [\psi]$. Furthermore, $z$ is dimensionless and $z \in \mathds{R}$.

Assuming the generalizing profile \eqref{Eq43} together with the scalar field solution $\psi$ announced in Eq. \eqref{Eq31}, the function $\mathcal{F}(\psi)$ takes the form
\begin{align}\label{Eq44}
	\mathcal{F}(\psi; r)=	\text{sinh}[q\,\nu^2\,\tanh[\eta\,\text{ln}(r)]^2]^{-\frac{1}{2z+1}}.
\end{align}
Consequently, for the generalizing function \eqref{Eq44}, the scalar field superpotential \eqref{Eq29}, and the BPS property constraint ($\tilde{W} = \cos^2 f$), it follows that the interaction \eqref{Eq11} responsible for spontaneous symmetry breaking in the topological sectors of the sigma and $\psi$ field is\footnote{In this case, one highlights the omnipresence of the breaking of translational symmetry. Thus, this breaking indicates to us that our system is no longer invariant under spatial displacements, which can give rise to novel and intriguing physical behaviors. For instance, when a system loses translational symmetry, it may lead to the emergence of new phases or deformations in the field variables, as well as the appearance of states that would not be possible in a system with preserved translational symmetry. This phenomenon plays a significant role in condensed matter physics, where the breaking of translational symmetry is often associated with phase transitions, such as the transition from a superconducting to a normal state or the formation of a non-periodic crystalline phase \cite{Kittel,Ashcroft}.}
\begin{align}\label{Eq45}
	V(f; r)=\cos f+\frac{\lambda \nu^4}{2r^2}\text{sech}[\eta\,\text{ln}(r)]^4.
\end{align}

In this framework, the magnetic vortices radiate an energy called BPS energy. Considering Eq. \eqref{Eq28}, it follows that the magnetic vortex energy in natural units is
\begin{align}\label{Eq46}
    \text{E}_{\text{BPS}}=1+N(\beta-1),
\end{align}
whose corresponding BPS energy density is
\begin{align}\label{Eq47}
    \rho_{\text{BPS}}=f'\,^2\mathcal{F}\,^{1/2}-\sqrt{2}\,\cos^2f+\frac{W_\psi^2}{r}.
\end{align}
 %Naturalmente, esta energia irradiada deve-se ao fluxo eletromagnetico que estas estruturas topológicas emitem\footnote{O fluxo de radiação magnética está exposto na Eq. \eqref{Eq26}.}.

To conclude our analysis, we need to solve the equations 
\begin{align}\label{Eq48}
	f'(r)=\pm\frac{N}{r}\text{sinh}[q\,\nu^2\,\tanh[\eta\,\text{ln}(r)]^2]^{\frac{1}{2(2z+1)}}(a-1)\sin f
\end{align}
and
\begin{align}\label{Eq49}
a'(r)=\pm\frac{\sqrt{2}r}{N}\cos f.
\end{align}
These expressions are known as the BPS vortex equations from the sigma model. By a quick inspection, one notes the absence of analytical solutions for Eqs. \eqref{Eq48} and \eqref{Eq49}. Thus, to investigate these solutions, let us adopt the numerical method outlined in Sec. \ref{sec2d1}..

\begin{figure}[!ht]
    \centering
    \subfigure[]{\includegraphics[width=8cm,height=7cm]{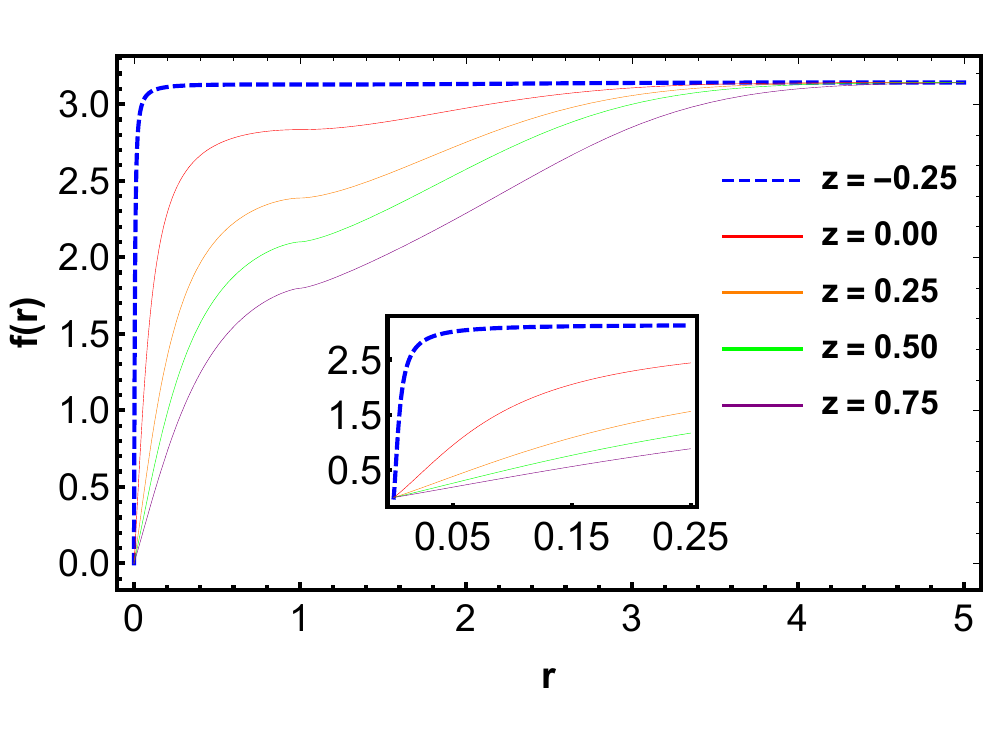}}\hfill
    \subfigure[]{\includegraphics[width=8cm,height=7cm]{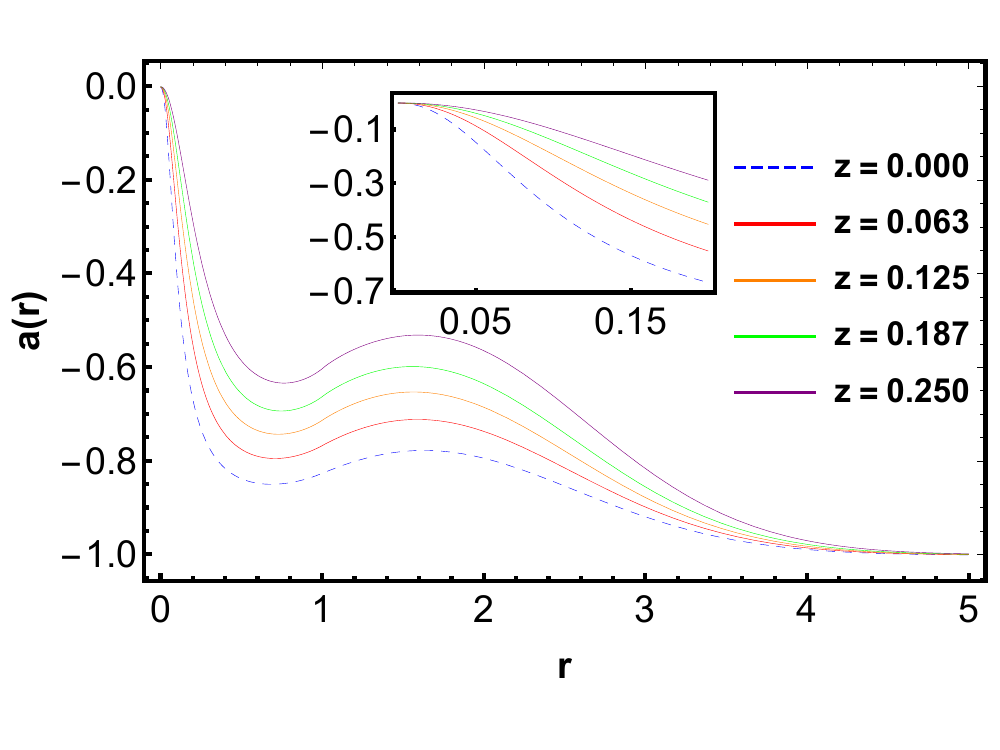}}
    \caption{Numerical solutions for the field profiles $f(r)$ [figure (a)] and $a(r)$ [figure (b)] vs. $r$ with $N=\lambda=\nu=e=1$.}
    \label{Fig2}
\end{figure}

By adopting the numerical approach (see Sec. \ref{sec2d1}) together with the boundary conditions \eqref{Eq24}, we find the solutions for the field variables of the O(3)-sigma model, $f$, and the gauge field, $a$. One exposes these solutions in Figs. \ref{Fig2}(a) and \ref{Fig2}(b), respectively.

When inspecting the numerical solutions of the field variables $f$ and $a$, our results indicate that, in the range $z \in [-0.25,0]$, the solutions exhibit a kink-like profile in the sigma-field sector. These profiles continuously deform into profiles resembling double-kink configurations. Consequently, the gauge field variable displays a global minimum at the vortex core ($r=0$), decaying for $a(r \to \infty) \to \beta=-1$, while undergoing a small oscillation (a local minimum followed by a local maximum) before reaching its plateau. We showed these behaviors in Figs. \ref{Fig2}(a) and \ref{Fig2}(b).

\begin{figure}[!ht]
  \centering
  \includegraphics[width=8cm,height=7cm]{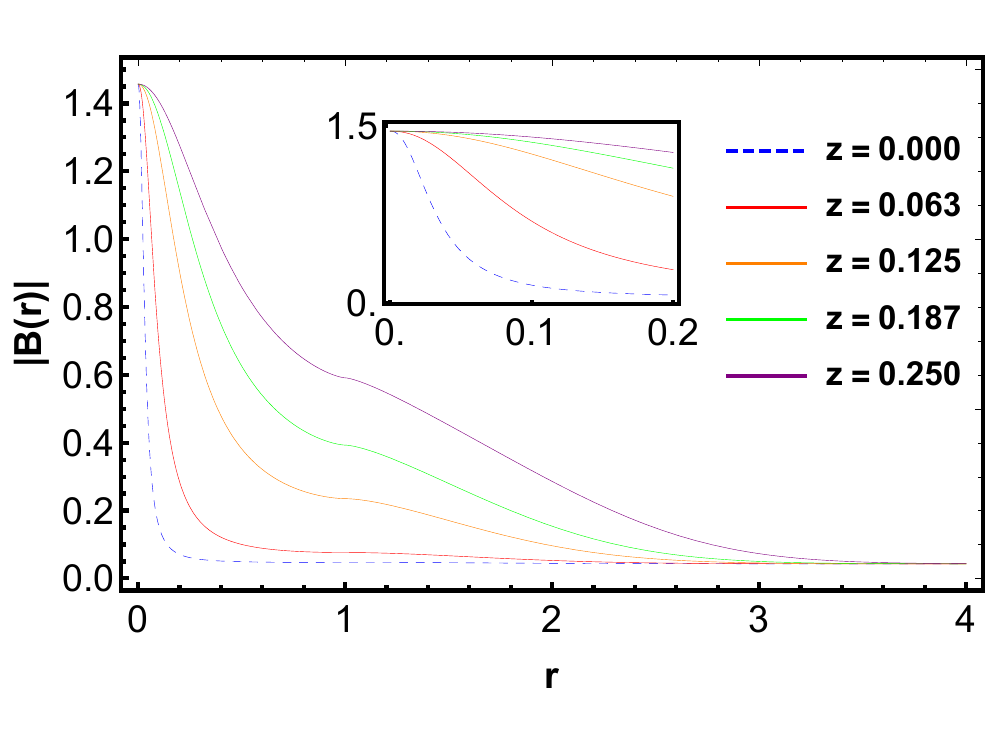}\\
  \subfigure[Case $z=0.000$.]{\includegraphics[height=4cm,width=4cm]{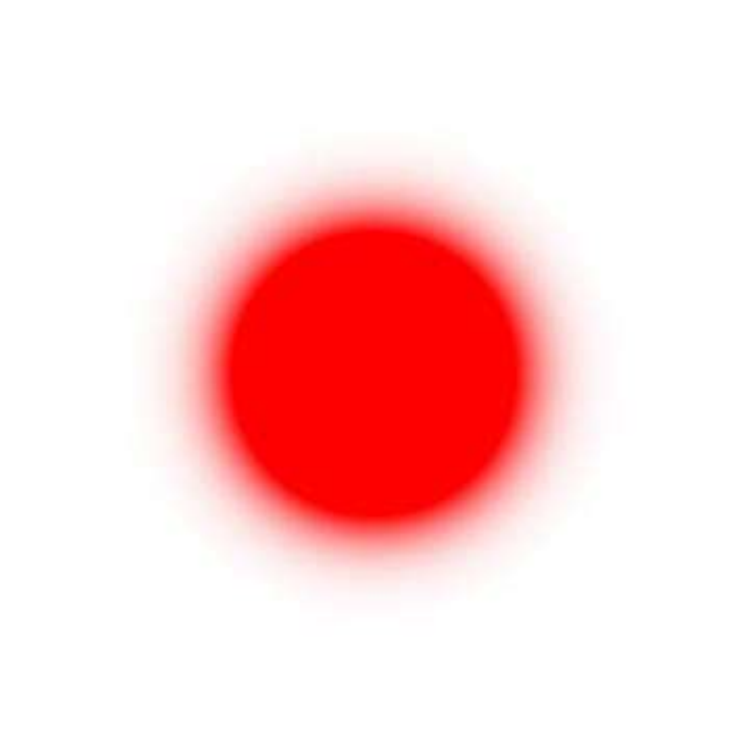}}\hfill
   \subfigure[Case $z=0.063$.]{\includegraphics[height=4cm,width=4cm]{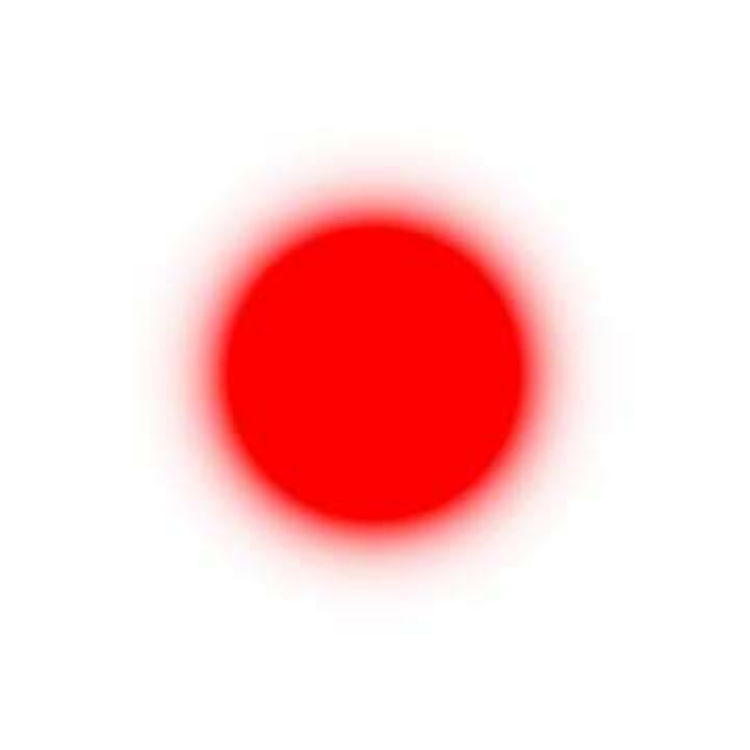}}\hfill
   \subfigure[Case $z=0.125$.]{\includegraphics[height=4cm,width=4cm]{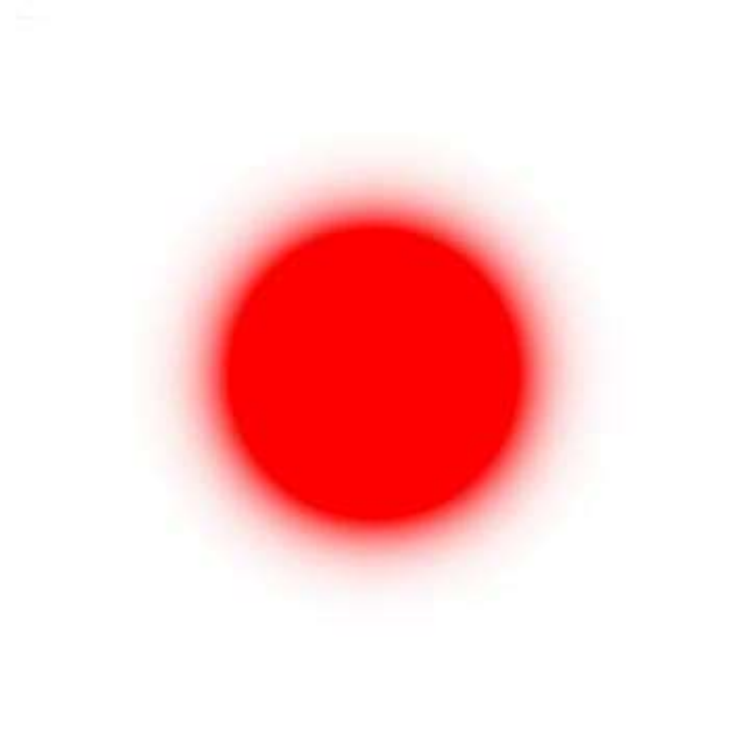}}\hfill
   \subfigure[Case $z=0.187$.]{\includegraphics[height=4cm,width=4cm]{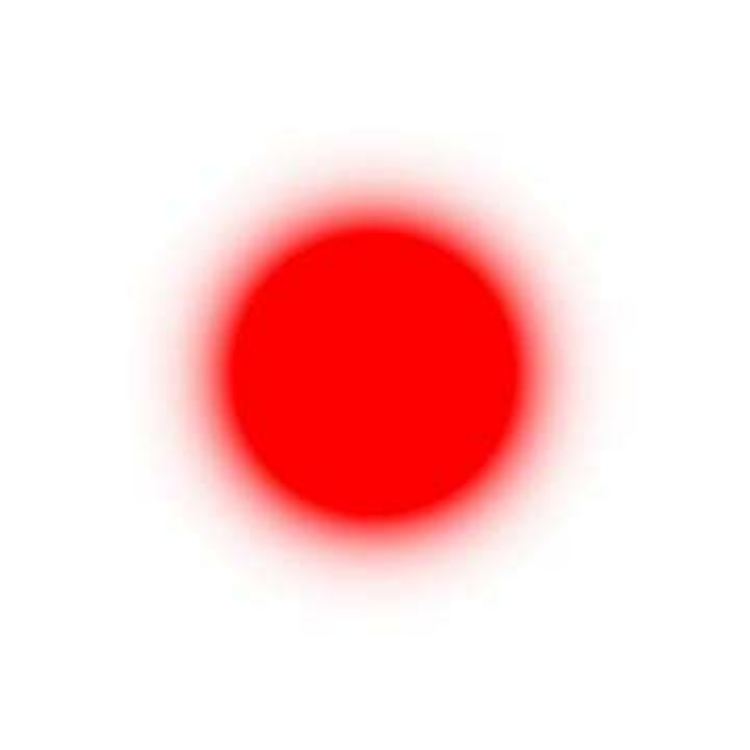}}
  \caption{Vortex magnetic field $\vert B(r)\vert$ vs. $r$ with $N=\lambda=\nu=e=1$.}
  \label{Fig3}
\end{figure}

Naturally, the behavior of the sigma [Fig. \ref{Fig2}(a)] and gauge field [Fig. \ref{Fig2}(b)] together with the scalar field [Eq. \eqref{Eq31}] gives rise to disk-shaped magnetic vortex profiles (see Fig. \ref{Fig3}). However, when the parameter $z$ increases, the vortex expands in size while maintaining its radiated magnetic flux, i.e., $\varphi_{\text{flux}} = 2\pi \beta$, and its energy, $\text{E}_{\text{BPS}} = \beta$ \footnote{The magnetic field and the vortex energy take these values once the winding number is $N=1$. Furthermore, we adopted the natural units, i.e., $\hbar=c=e=1$.}.

\begin{figure}[!ht]
  \centering
  \includegraphics[width=8cm,height=7cm]{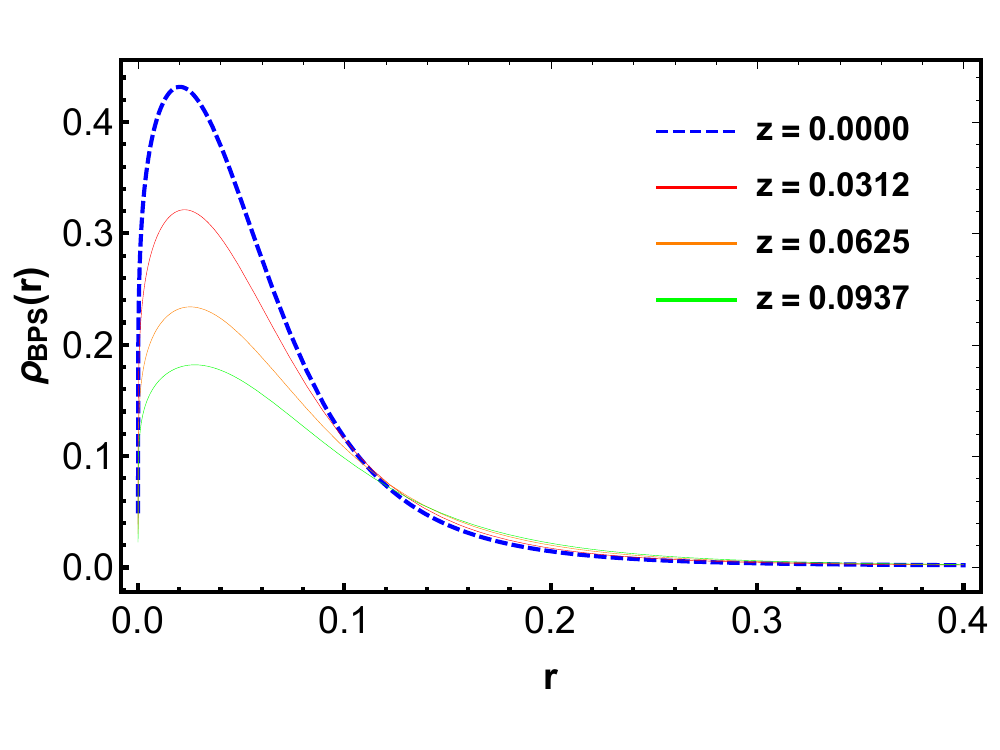}\\
  \subfigure[Case $z=0.0000$.]{\includegraphics[height=4cm,width=4cm]{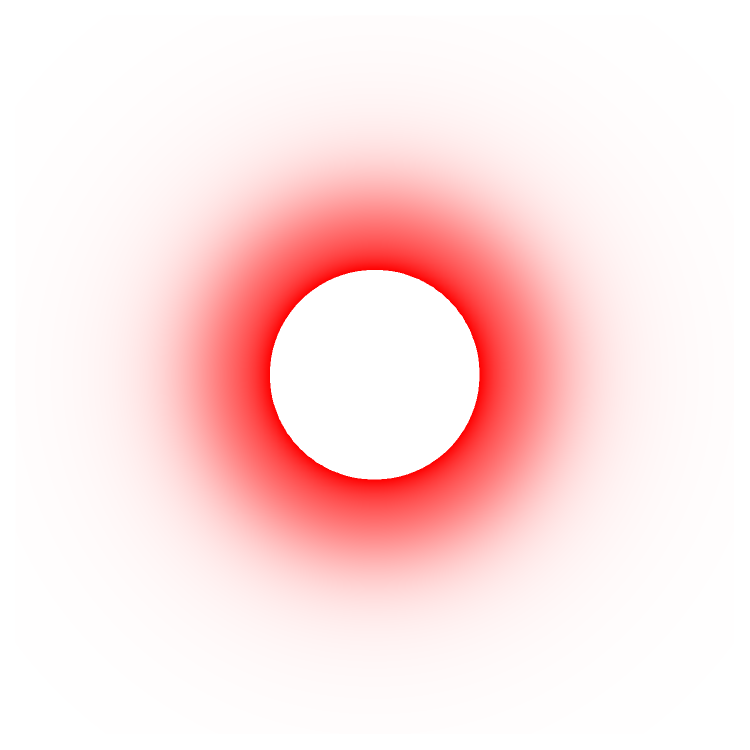}}\hfill
   \subfigure[Case $z=0.0312$.]{\includegraphics[height=4cm,width=4cm]{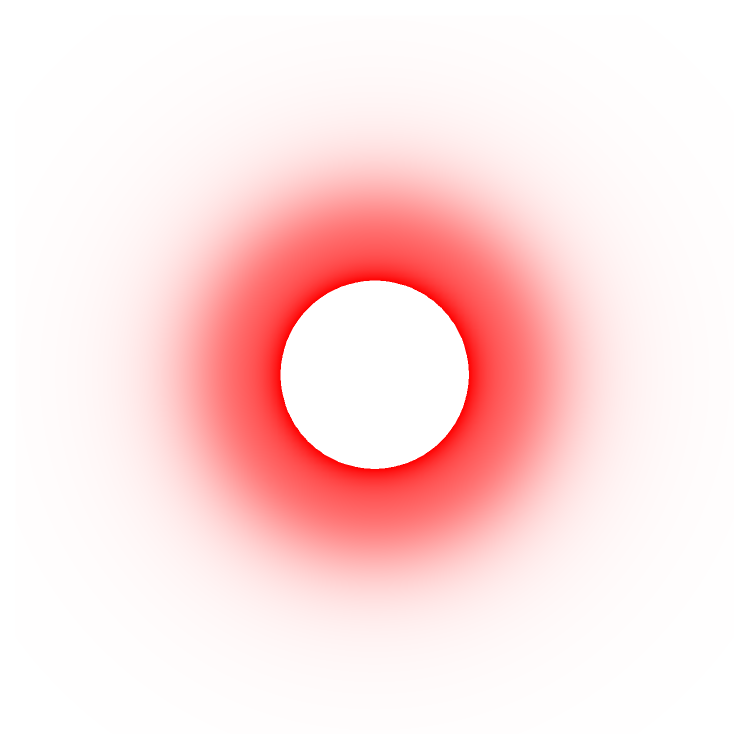}}\hfill
   \subfigure[Case $z=0.0625$.]{\includegraphics[height=4cm,width=4cm]{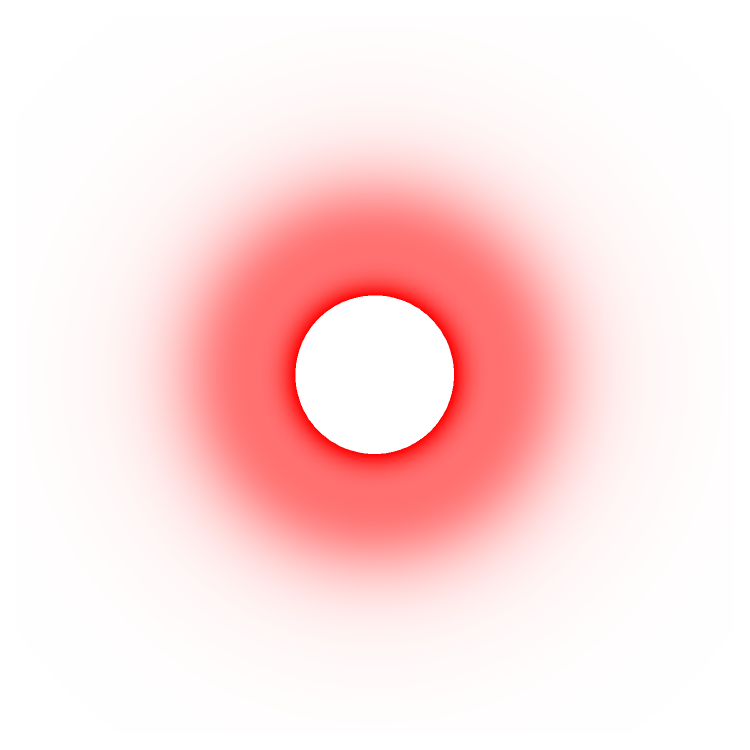}}\hfill
   \subfigure[Case $z=0.0937$.]{\includegraphics[height=4cm,width=4cm]{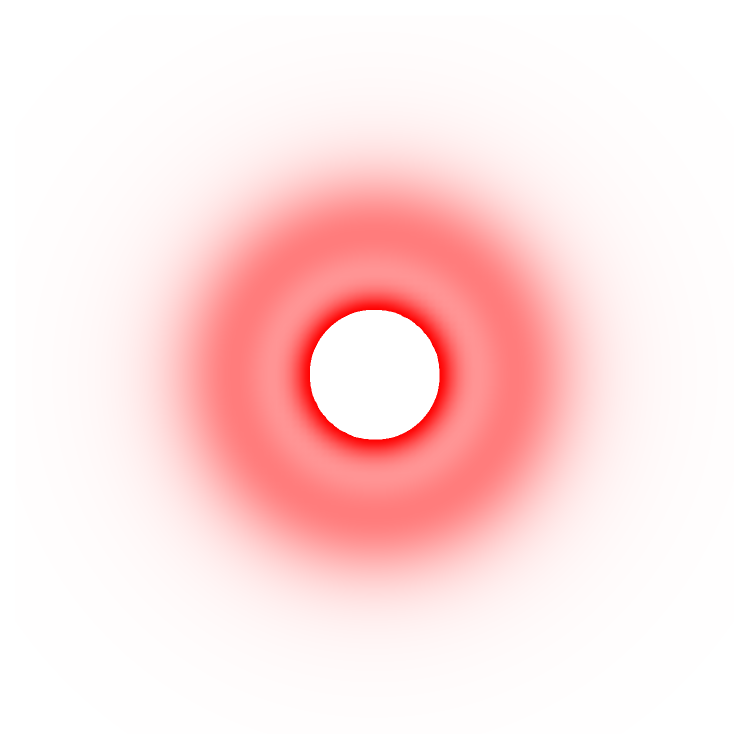}}
  \caption{BPS energy density $\text{E}_{\text{BPS}}$ vs. $r$ when $N=\lambda=\nu=e=1$.}
  \label{Fig4}
\end{figure}

To analyze the magnetic field and the energy from the vortex, we consider the numerical solutions [Figs. \ref{Fig2}(a) and \ref{Fig2}(b)], together with the numerical approach announced in Sect. \ref{sec2d1}, the definition of the magnetic field, $\vert B\vert = -F_{21} = \frac{N a}{e\,r}$ \cite{Weinberg}, and the BPS energy density [Eq. \eqref{Eq47}]. This approach yields the numerical results for the magnetic field (see Fig. \ref{Fig3}) and the energy density (see Fig. \ref{Fig4}).

Examining the behavior of the magnetic field (Fig. \ref{Fig3}), one notes that as the deformation parameter of the generalizing function, $z$, increases, it alters the field variables $f$ and $a$. Consequently, this induces an increase (or vortex size adjustment) in the radial size of the magnetic vortex (see Fig. \ref{Fig3}), while these disk-like vortices maintain a total magnetic flux given by $\varphi_{\text{flux}} = 2\pi \beta$, resulting in a geometric expansion of the magnetic vortices. Meanwhile, the energy profiles (Fig. \ref{Fig4}) have a ring-like profile. These energetic rings decrease asymmetrically around the point of maximum energy (near the core), remaining localized around the core. One highlights that although the energy density is attenuated near the vortex core (with energy density null at the $r=0$), the total energy remains constant, independent of variations of the generalizing function parameter $z$, which controls the deformation of the structures.

\subsubsection{Behavior of the solutions and their phenomenological connection}

Finally, to conclude our study, let us analyze the asymptotic behavior of the field variables $f(r)$ and $a(r)$ when $r\to 0$. For simplicity of argument, we consider the case whose winding number is positive, i.e., $N=m>0$. In this framework, one notes that, in the limit $r \to 0$, the field variables described by Eqs. \eqref{Eq47} and \eqref{Eq48} take the following form
\begin{align}\label{Eq50}
    a(r)\approx C_m r^2 \hspace{1cm} \text{and} \hspace{1cm}
    f(r)\approx \overline{C}_m r,
\end{align}
where $C_m=-\frac{\sqrt{2}}{2m}$ and $\overline{C}_{m}=\frac{2^{9/4}\text{e}^{-\frac{(m-1)^2}{\sqrt{2}}}}{\sqrt{\pi}\,\text{erf}\left(\frac{m-1}{2^{1/4}}\right)}$ \footnote{Remember that $\lambda=\nu=e=1$ and $\eta=\nu\sqrt{\lambda}=1$.}. 

Once the behavior of the field variables $a(r)$ and $f(r)$ when $r\to 0$ is known, one can conclude that the BPS energy density near the vortex center is
\begin{align}
    \rho_{\text{BPS}}(r)\approx-\frac{2\,\text{sech}[\,\text{ln}(r)]^2\tanh[\,\text{ln}(r)]}{r}.
\end{align}
Meanwhile, the magnetic field boils down to
\begin{align}
    \vert B(r)\vert\approx\frac{\sqrt{2}}{2}\left(2-\overline{C}\,_{m}^{2}r^2\right).
\end{align}
Thus, it is notorious that the above expressions are in agreement with the numerical results presented in the previous section (see Figs. \ref{Fig2}, \ref{Fig3}, and \ref{Fig4}).

To summarize, let us examine the asymptotic behavior of the system, i.e., when $r \to \infty$. Again, considering the expressions \eqref{Eq47} and \eqref{Eq48}, one notes that, when $r\to\infty$, the solutions approach constant values, viz.,
\begin{align}
    &a(r)\approx -\beta \hspace{1cm} \text{and} \hspace{1cm} f(r)\approx\pi, %\overline{M}\,_{N,q,\nu}\,\text{ln}\left(\frac{r}{r_0}\right).
\end{align}
This results in a vanishing BPS energy density ($\rho_{\text{BPS}} = 0$) and a zero magnetic field (i.e., $\vert B \vert \approx 0$, considering the magnetic flux produced by these structures). Naturally, this confirms that the magnetic field is maximal at the vortex core. However, as the radial distance increases toward infinity, the magnetic field tends to the constant $\beta = 0$. Thus, at far-reaching distances, the magnetic field becomes negligible. Nevertheless, the magnetic flux of these vortices remains constant and quantized, i.e., $\varphi_{\text{flux}} = \frac{2\pi N \beta}{e}$, with the scalar, sigma, and gauge fields approaching their respective vacuum values.

Consequently, the magnetic vortices obtained here resemble Abrikosov-like vortices (vortices in type-II superconductors) \cite{Abrikosov}, as they represent magnetic vortices whose energy of a single vortex is lower than the total magnetic flux sustaining the emergent magnetic field. Furthermore, one notes that, similarly to Abrikosov vortices in type-II superconductors, the vortices in this theory also maintain a regular singularity within their configuration, i.e., at the core ($r=0$). It is essential to highlight that vortices with such characteristics can, in principle, be trapped by topological defects in the superconducting material, influencing its behavior, including the destruction of the superconducting state and the confinement of magnetic flux. Therefore, these structures are of great interest for technological applications, such as in the development of magnets \cite{Wilson}, superconductors, and superconducting quantum interference devices (SQUIDs) \cite{Kleiner}, with direct applications in electronic devices, including in the medical branch \cite{Hari}.

\section{Summary and conclusions}

Throughout this work, we investigated the formation and geometric expansion of ANO-like magnetic vortices within a noncanonical multi-field theory. To accomplish our purpose, one adopts a framework whose the noncanonical O(3)-sigma model couples nonminimally to a gauge field and a real scalar field. We justified this theoretical choice due to its ability to describe topological structures that admit BPS property and sustain a quantized magnetic flux, a phenomenon essential for understanding magnetic vortices.

We developed this formulation using the BPS approach, which enabled the derivation of self-dual equations ensuring the stability and minimum energy configurations. One notes that, implementing a noncanonical hyperbolic extension, the vortex solutions undergo a geometric expansion, acquiring wider disk-like profiles. This deformation notably generates concentric energy rings around the vortex core, mitigating energy concentration while preserving both flux quantization and the constancy of the total system energy.

The analysis of the solutions required a numerical approach. Thus, when adopting the Runge-Kutta approach, one obtains the solutions of the coupled differential equations and a detailed characterization of the magnetic vortex configurations. The results show that the profiles of the sigma, the scalar, and the gauge fields vary with the introduced deformation parameter, without compromising the fundamental topological properties that guarantee solution stability.

To summarize, the proposed model not only confirms the existence of noncanonical BPS vortices with quantized magnetic flux and stable energy but also reveals the possibility of controlled geometric expansion of these structures. This phenomenon brings the studied vortices closer to Abrikosov vortices in type-II superconductors, since both exhibit a singularity at the core and a single-vortex formation energy lower than the total sustained magnetic flux. Taking these aspects into account, one notes that our results increase the effectiveness of the theoretical model and its technological applications, particularly when applied to superconducting materials, the development of magnetic flux confinement devices, and highly sensitive instruments such as SQUIDs, widely employed in medical physics and in technologies for detecting extremely weak magnetic fields.

\section*{ACKNOWLEDGMENT}
The authors would like to express their sincere gratitude to the Conselho Nacional de Desenvolvimento Científico e Tecnológico (CNPq) and Fundação de Amparo \`{a} Pesquisa do Estado de S\~{a}o Paulo (FAPESP) for their valuable support. F. C. E. Lima is supported, respectively, for grants No. 2025/05176-7 (FAPESP) and 171048/2023-7 (CNPq). 

The author expresses his sincere gratitude to Prof. D. Vassilevich for the valuable and fruitful discussions that greatly contributed to the development of this work.

\section*{DATA AVAILABILITY}

No data was used for the research described in this article.

\end{document}